\documentclass{article}
\PassOptionsToPackage{numbers, compress}{natbib}
\usepackage{arxiv}
\usepackage{natbib}
\providecommand{\IEEEmembership}[1]{\textit{#1}}

\usepackage[utf8]{inputenc}
\usepackage[T1]{fontenc}
\usepackage{hyperref}
\usepackage{url}
\usepackage{booktabs}
\usepackage{amsfonts}
\usepackage{amssymb}
\usepackage{amsmath}
\usepackage{amsthm}
\usepackage{graphicx}
\graphicspath{{./}{../}}
\usepackage[table]{xcolor}
\usepackage{algorithm}
\usepackage{algorithmic}
\usepackage{microtype}
\usepackage{multirow}
\usepackage{wrapfig}
\usepackage{subfig}
\usepackage[capitalise]{cleveref}
\usepackage{enumitem}
\Crefname{section}{Sec.}{Secs.}
\Crefname{subsection}{Sec.}{Secs.}
\Crefname{subsubsection}{Sec.}{Secs.}
\Crefname{Equation}{Equation}{Equations}
\Crefname{Figure}{Figure}{Figures}
\Crefname{Table}{Table}{Tables}

\theoremstyle{definition}     

\newcommand{\bR}{\mathbb{R}}

\newcommand{\bI}{\mathbf{I}}
\newcommand{\xhat}{\widehat{\mathbf{x}}_0}
\newcommand{\calL}{\mathcal{L}}
\newcommand{\bmu}{\boldsymbol{\mu}}
\newcommand{\bdelta}{\boldsymbol{\delta}}

\title{DUET: Unified Dual-Space Emotion Control for Diffusion and Flow-Matching Driven Text-to-Speech}

\author{%
Xu~Zhang,
Longbing~Cao\thanks{Corresponding author. Xu Zhang, Longbing Cao, and Zhangkai Wu are with the Frontier AI Research Centre, Macquarie University, Sydney, NSW, Australia (e-mail: xu.zhang12@hdr.mq.edu.au; longbing.cao@mq.edu.au; zhangkai.wu@mq.edu.au).},~\IEEEmembership{Senior Member,~IEEE},
and~Zhangkai~Wu%
}

\begin{document}

\date{}
\maketitle

\begin{abstract}
Diffusion and flow-matching based text-to-speech (TTS) models excel in naturalness but often lack explicit emotion control, as emotional signals remain entangled with speaker identity. We discover that emotion embedding emerges as a linearly decodable direction of frozen hidden states, nearly orthogonal to the direction embedding speaker identity. This inspires a plug-and-play framework DUET for emotion control over pretrained diffusion and flow-matching based TTS models. During generation, DUET unifies dual-space control to achieve fine-grained emotion intervention in a single per-step update: \textit{hidden space steering} shifts generation along the target emotion direction, while \textit{mel-space guidance} refines spectral details through gradients backpropagated from a differentiable vocoder. We validate DUET on five architecturally diverse pretrained TTS backbones across three datasets, where it outperforms 10 supervised state-of-the-art emotional TTS baselines across paradigms and achieves the highest human-rated emotion appropriateness. To further showcase its qualitative behavior, we deploy DUET on an Ameca humanoid robot, where it produces richly expressive emotional speech on the humanoid, demonstrating the strong potential for plug-and-play affective interaction for embodied agents.
\end{abstract}

\section{Introduction}
\label{sec:intro}

    Recent iterative generative Text-to-Speech (TTS) models have substantially improved speech naturalness, where diffusion and flow matching models provide two dominant formulations for progressive spectrum generation \citep{chen2025f5tts,popov2021gradtts,mehta2024matcha,huang2022prodiff,wu2025sepdiff,wu2025progdiffusion}. By iteratively denoising mel spectrograms, these models capture the expressive distribution of natural speech, reproducing both fine acoustic textures and prosodic diversity. Emotional variants \citep{guo2023emodiff,cho2024emosphere,cho2025diemo} extend these backbones with emotion conditioning by retraining each base model on large emotion-labeled corpora.

\begin{figure}[t]
  \centering
  \subfloat[Hidden state variance.\label{fig:discovery_a}]{%
    \includegraphics[width=0.235\textwidth]{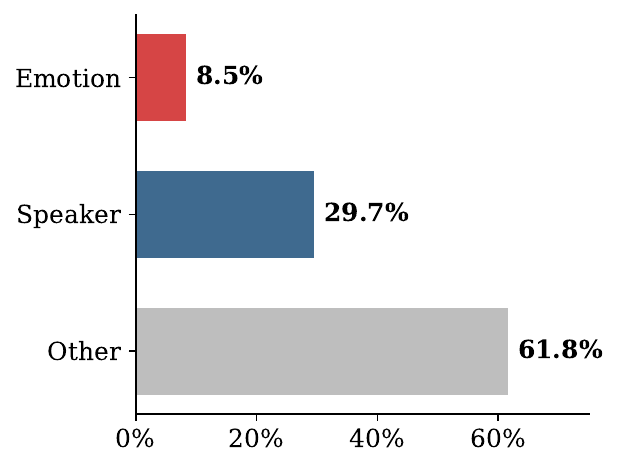}}\hfill
  \subfloat[Cross speaker cosine.\label{fig:discovery_b}]{%
    \includegraphics[width=0.235\textwidth]{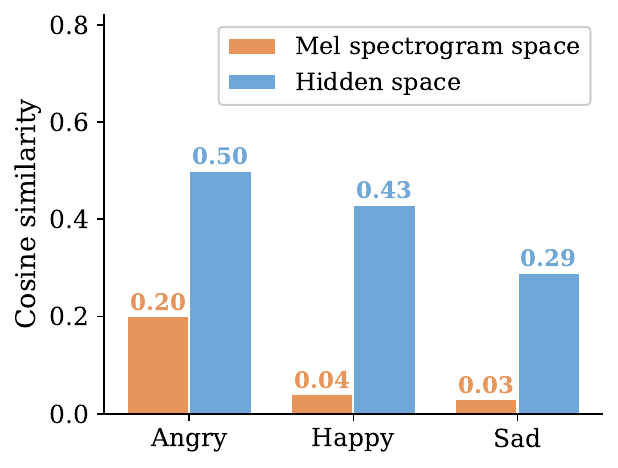}}\hfill
  \subfloat[Hidden state clusters.\label{fig:discovery_c}]{%
    \includegraphics[width=0.235\textwidth]{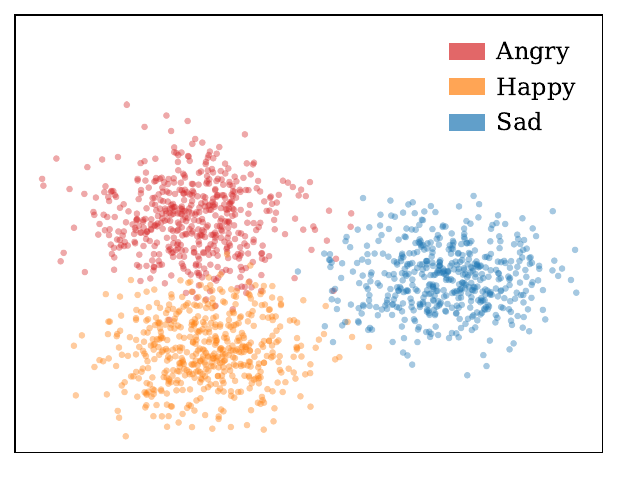}}\hfill
  \subfloat[Layer probe accuracy.\label{fig:discovery_d}]{%
    \includegraphics[width=0.235\textwidth]{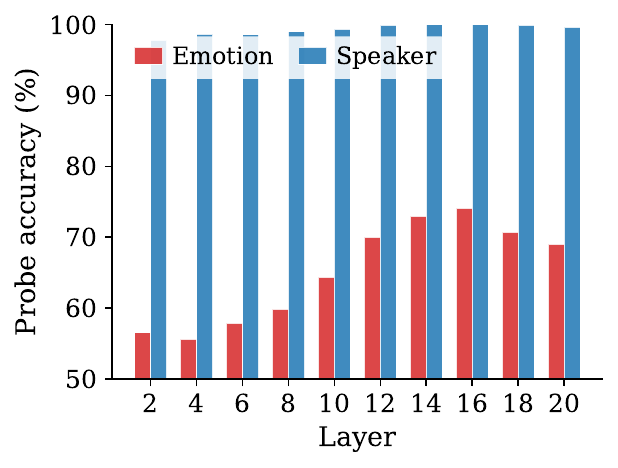}}
  \caption{Emotion is subtle but speaker-consistent and linearly decodable in pretrained TTS hidden states. \textbf{(a)} Emotion accounts for only 8.5\% of hidden state variance, dominated by speaker identity and other details; \textbf{(b)} Cross-speaker cosine similarity of same emotion representations is low in mel-space but substantially higher in hidden space, identifying hidden states as the speaker-consistent representation; \textbf{(c)} Hidden states at layer 16 of a frozen TTS model form well separated emotion clusters despite no emotion supervision; \textbf{(d)} Linear probe accuracy across layers shows emotion separability peaking at middle layers while speaker identity remains uniformly decodable.}
  \label{fig:discovery}
  \vspace{-12pt}
\end{figure}

    Emotion control over pretrained TTS models is promising because it avoids costly supervised retraining, while enabling direct reuse of effective TTS-oriented diffusion and flow-matching backbones. However, precise emotion control in these models remains challenging because emotion is a subtle factor in hidden space, accounting for only a small portion of the representation variance (\Cref{fig:discovery}(a)). Its acoustic realization is also closely coupled with speaker identity, so the same emotion appears differently across speakers in the mel-space (\Cref{fig:discovery}(b)). This motivates us to examine whether emotion in frozen TTS models admits a ``speaker-coherent geometric representation''. Our analysis reveals that \textit{emotion embedding emerges as a linearly decodable directions in frozen hidden states, nearly orthogonal to the direction embedding speaker identity}. We discover that hidden states form emotion-aware clusters without emotion supervision (\Cref{fig:discovery}(c)), and linear probing shows that emotion separability peaks in the middle layers while speaker identity remains consistently decodable across layers (\Cref{fig:discovery}(d)).

    These findings motivate unified \underline{\textbf{DU}}al-space \underline{\textbf{E}}motion con\underline{\textbf{T}}rol (\textbf{DUET}), a plug-and-play framework that steers hidden states along the direction embedding target emotion while refining mel-space details during denoising, instead of retraining with a fixed emotion taxonomy. Specifically, DUET steers the hidden state at the most discriminative layer along a target emotion direction extracted by a probing procedure, biasing the denoising prediction toward the target emotion. Moreover, to reach the fine spectral details that hidden state steering alone cannot affect, DUET refines the denoised intermediate mel by backpropagating the gradient of an external emotion recognizer through a differentiable vocoder. The two complementary interventions are unified into a single per-step update to shape the global prosodic trajectory and recover fine acoustic textures of the target emotion.

    Extensive evaluation of DUET across three datasets and five architecturally diverse backbones validates its plug-and-play generality. Its superior performance over 10 supervised emotional TTS baselines across paradigms and the highest emotion appropriateness ratings in human evaluation verify the effectiveness of our emotion control. Our contributions can be summarized as:
  \begin{itemize}
    \item We show that emotion in the hidden states of pretrained diffusion and flow-matching based TTS models is linearly decodable and nearly orthogonal to those encoding speaker identity, opening a new path toward precise training-free emotion control on frozen backbones.
    \item We propose DUET, a dual-space framework that unifies hidden state steering along a target emotion direction with mel-space guidance via gradient through a differentiable vocoder, enabling fine-grained emotion control on diffusion and flow-matching based TTS models.
    \item Extensive cross-architecture evaluations validate the plug-and-play generality and effectiveness of DUET across various diffusion and flow-matching based TTS backbones. 
    \item We further deploy DUET on an Ameca humanoid robot, where it produces richly expressive emotional speech for embodied interaction. This further demonstrates the strong potential of DUET-enabled affective interaction for embodied agents.
  \end{itemize}
  
\section{Iterative Generative TTS Models}
\label{sec:prelim}
    The pretrained iterative generative TTS framework contains two parts: speech embedding construction and iterative spectrum generation. Given a text $\mathbf{u}=\{u_i\}_{i=1}^{L}$ and an optional reference utterance $\mathbf{r}_{\mathrm{ref}}$, the encoder maps the inputs to a frame-level speech embedding $\mathbf{c}=\psi(\mathbf{u},\mathbf{r}_{\mathrm{ref}})$ that carries linguistic content and speaker information. Conditioned on $\mathbf{c}$, the iterative spectrum generator produces a mel spectrogram $\mathbf{x}_0\in\bR^{F\times M}$, where $F$ is the number of acoustic frames and $M$ the mel dimension. A separate vocoder $g_{\mathrm{voc}}$ renders the spectrogram into a waveform $\mathbf{y}=g_{\mathrm{voc}}(\mathbf{x}_0)$, and its differentiability allows acoustic gradients to flow back to $\mathbf{x}_0$ during sampling.
    
    The iterative spectrum generator can be either diffusion or flow-matching model. For diffusion-based TTS, the acoustic state $\mathbf{x}_t\in\bR^{F\times M}$ follows a forward SDE:
$
    \mathrm{d}\mathbf{x}_t
    =
    f(\mathbf{x}_t,t;\,\mathbf{c})\,\mathrm{d}t
    +
    g(t)\,\mathrm{d}\mathbf{w}_t,
    t\in[0,1],
$
where $\mathbf{x}_0$ is the clean mel and $\mathbf{x}_1$ is close to Gaussian noise. Sampling integrates the reverse SDE:
$
    \mathrm{d}\mathbf{x}_t
    =
    \big[
    f(\mathbf{x}_t,t;\,\mathbf{c})
    -
    g(t)^2 s_\theta(\mathbf{x}_t,t;\,\mathbf{c})
    \big]\mathrm{d}t
    +
    g(t)\,\mathrm{d}\bar{\mathbf{w}}_t,
$
where $s_\theta$ is the score of the pretrained model and $\bar{\mathbf{w}}_t$ is the reverse-time Brownian motion. For flow-matching TTS, the trajectory follows the probability flow ODE \citep{wu2025scot}:
$
    \mathrm{d}\mathbf{x}_t
    =
    v_\theta(\mathbf{x}_t,t;\,\mathbf{c})\,\mathrm{d}t,
$
with velocity field $v_\theta$. At each sampling step, both predictors admit a one-step extrapolation to a clean mel estimate:
\begin{equation}
\label{eq:clean_estimate}
\xhat =
\begin{cases}
    \mathbf{x}_t - t\,v_\theta(\mathbf{x}_t,t;\,\mathbf{c}),
    & \text{flow matching}, \\[4pt]
    \big(
    \mathbf{x}_t
    -
    \sqrt{1-\bar\alpha_t}\,
    \epsilon_\theta(\mathbf{x}_t,t;\,\mathbf{c})
    \big)
/ \sqrt{\bar\alpha_t},
    & \text{diffusion},
\end{cases}
\end{equation}
    where the noise predictor $\epsilon_\theta=-\sqrt{1-\bar\alpha_t}\,s_\theta$ is the standard reparameterization of the score and $\bar\alpha_t=\prod_{\tau\le t}\alpha_\tau$ is the cumulative noise schedule \citep{ho2020ddpm}. 
    
    Iterative spectrum generation therefore exposes two intermediates at every step: the clean estimate $\xhat$ from \Cref{eq:clean_estimate}, and hidden states $\mathbf{h}_l\in\bR^{D_l\times S_l}$ at every layer $l$ of the denoiser, where $D_l$ is the hidden dimension and $S_l$ is the sequence length. We aim to control these pretrained TTS models to generate speech in a target emotion $e \in \{\text{happy}, \text{angry}, \text{sad} \}$. However, without explicit emotion supervision during pretraining, emotion is entangled across emotion classes and with speaker identity in $\mathbf{h}_{l}$. Furthermore, subtle emotional adjustments to $\xhat$ create a mismatch with $g_{\mathrm{voc}}$ and fail to appear faithfully in the waveform. Together, these leave the model without precise emotion controllability.

\section{DUET: Unified Dual-Space Emotion Steering}
\label{sec:method}
    DUET builds on the insight that hidden states admit a speaker-coherent geometry for emotion, providing a representational interface for steering, while the clean mel estimate provides a mel-space interface for guidance aligned with $g_{\mathrm{voc}}$. DUET proceeds in three stages: (1) hidden state probing identifies the most emotion-discriminative layer and discriminant direction $\mathbf{d}_e$; (2) hidden state steering shifts $\mathbf{h}_{l}$ along $\mathbf{d}_e$ toward the target emotion; and (3) mel-space guidance corrects the clean mel estimate with gradients backpropagated through a differentiable vocoder. Finally, we integrate these two interventions on $\mathbf{h}_{l}$ and $\xhat$ into diffusion and flow-matching samplers.

\subsection{Hidden State Steering via Discriminant Directions}
\label{sec:latent}
    To extract the discriminant direction $\mathbf{d}_e$ that DUET applies at each denoising step, we first identify the emotion separable layer in \Cref{sec:probhidden} and then compute an emotion direction in \Cref{sec:disdirections}.
    
\subsubsection{Probing Hidden States for Emotion}
\label{sec:probhidden}
    We design a probing procedure to identify emotion discriminant directions and verify their disentanglement from speaker identity. At each layer $l$ of a frozen iterative generative TTS model, we attach a lightweight linear probe $\phi_l(\mathbf{h}_l) = \mathbf{W}_l\mathbf{h}_l + \mathbf{b}_l$ with utterance-level logits obtained by mean pooling across frames, trained via cross-entropy on emotion labels \citep{li2023iti,todd2024function,wu2024reft}. The layer with highest emotion recognition accuracy is selected as $l^e$, and its weight matrix $\mathbf{W}_{l^e}$ supplies the discriminant directions, providing both the layer and direction needed for steering the hidden state.

    However, the directions derived from $\mathbf{W}_{l^e}$ may be entangled with speaker identity, since hidden states encode emotion and speaker through overlapping prosodic cues such as pitch, energy, and timing. To test whether such a separation exists, we employ another speaker probe and measure the cosine similarity between the speaker and emotion directions. We denote this shared probing layer as $l^{e,s}$, where $s$ is speaker identity. Our analysis shows that two directions occupy nearly orthogonal subspaces across the backbones. For example, on F5-TTS, $|\cos\theta|=0.029$ at $l^{e,s}$. Steering along this direction therefore shifts the emotion direction while leaving the speaker identity direction nearly unchanged in the hidden state. Detailed similarity analysis is shown in \Cref{sec:repr_analysis}.

    Since backbones differ in how they encode emotion from external supervised information modality, we use two probing modes: \emph{reference speech probing} for backbones that condition on a reference speech, and \emph{generation time probing} for those that generate from text alone. In the former, we obtain training pairs $(\mathbf{h}_l, \mathbf{y})$ by passing labeled emotional reference speech through the frozen model and extracting $\mathbf{h}_l$ at each layer, with $\mathbf{y}$ the ground truth emotion label. For backbones that generate from text alone, reference speech probing does not transfer. Specifically, offline hidden states from reference speech are nearly orthogonal to those captured during generation, with cosine similarity around $0.05$ (similarity analysis across backbones in \Cref{sec:repr_analysis}). We therefore probe at generation time: we generate utterances from the frozen model, extract $\mathbf{h}_l$ at each layer, and label the synthesized speech with an external emotion recognizer to obtain pseudo labels $\hat{\mathbf{y}}$. The probe $\phi_l$ trained on $(\mathbf{h}_l, \hat{\mathbf{y}})$ pairs matches reference speech probing in accuracy. This labeling recognizer is disjoint from those used for evaluation. Across both backbones, the layer with highest emotion probe accuracy, where the emotion and speaker directions are nearly orthogonal, is denoted $l^*$ with probe weights $\mathbf{W}_{l^*}$.

\subsubsection{Discriminant Directions from Probe Weights}
\label{sec:disdirections}
    From $\mathbf{W}_{l^*}$, we construct a steering direction $\mathbf{d}_e$ that combines an emotion centroid with additional discriminative directions from the probe. To represent each emotion class by its mean in the hidden space, we average its hidden states over the probing data, defining $\bmu_e$ for target emotion $e$ and $\bmu_0$ for the neutral class. We define the centroid direction as $\bdelta_e = \bmu_e - \bmu_0$, which we normalize as $\hat{\bdelta}_e = \bdelta_e/\|\bdelta_e\|$, as in the activation steering \citep{turner2023actadd}. While $\bdelta_e$ represents only a single direction, it does not span the full discriminant directions defined by $\mathbf{W}_{l^*}$. We therefore extract $k$ additional directions from $\mathbf{W}_{l^*}$, which encodes the partition of hidden space into emotion classes, and aggregate them with $\bdelta_e$ to form the steering direction $\mathbf{d}_e$.

    To identify the most effective $k$ directions, we first set $\mathbf{W} = \mathbf{W}_{l^*}$ for brevity and evaluate each candidate unit vector $\mathbf{v} \in \bR^D$ by the probe sensitivity $\|\mathbf{W}\mathbf{v}\|^2$. Specifically, high sensitivity means small perturbations along $\mathbf{v}$ shift the probe output significantly, showing $\mathbf{v}$ as an effective steering direction. We select $k$ directions $\mathbf{v}_1, \dots, \mathbf{v}_k$ that maximize this sensitivity under the orthonormality of $\{\hat{\bdelta}_e, \mathbf{v}_1, \dots, \mathbf{v}_k\}$, formalized as:
\begin{equation}
  \label{eq:opt}
  \max_{\mathbf{v}_1, \dots, \mathbf{v}_k \in \bR^D}
  \sum_{i=1}^{k} \|\mathbf{W}\mathbf{v}_i\|^2.
\end{equation}
We obtain these directions by removing the component aligned with $\hat{\bdelta}_e$ from $\mathbf{W}$:
$
  \tilde{\mathbf{W}} = \mathbf{W}\,(\bI - \hat{\bdelta}_e\,\hat{\bdelta}_e^\top),
$
    where $\bI - \hat{\bdelta}_e\,\hat{\bdelta}_e^\top$ is the projector onto the orthogonal complement of $\hat{\bdelta}_e$, so any direction extracted from $\tilde{\mathbf{W}}$ is automatically orthogonal to $\hat{\bdelta}_e$. We then apply a singular value decomposition (SVD) to $\tilde{\mathbf{W}}$ and take its top-$k$ right singular vectors as $\mathbf{v}_1, \dots, \mathbf{v}_k$. These maximize the probe sensitivity and, together with $\hat{\bdelta}_e$, form the orthonormal set required by \Cref{eq:opt}. Finally, we form the steering direction:
\begin{equation}
      \mathbf{d}_e
  = \hat{\bdelta}_e
  + \beta \sum_{i=1}^{k} \mathbf{v}_i,
  \label{eq:direction}
\end{equation}
    where $\beta$ controls the relative weight of the probe-derived directions and the centroid, with each $\mathbf{v}_i$ sign-aligned to the target emotion class.

\subsubsection{Norm Adaptive Steering}
    When steering $\mathbf{h} = \mathbf{h}_{l^*}$ along the emotion direction $\mathbf{d}_e$ from \Cref{eq:direction}, the update scale should adapt to $\|\mathbf{h}\|$, which varies substantially across denoising steps. An unscaled perturbation would otherwise dominate $\mathbf{h}$ in early steps and become negligible relative to it in late steps. We scale the perturbation by $\|\mathbf{h}\|$ so its relative strength stays constant, yielding:
\begin{equation}
  \label{eq:latent_guidance}
  \mathbf{h}^{*}
  = \mathbf{h}
  + \lambda \cdot \frac{\mathbf{d}_e}{\|\mathbf{d}_e\|}
    \cdot \|\mathbf{h}\|\,,
\end{equation}
    where the update is taken per frame, with $\lambda$ the ratio of the per-frame perturbation norm to $\|\mathbf{h}\|$, kept constant across denoising steps. The resulting $\mathbf{h}^*$ propagates through the remaining layers and steers the velocity prediction for flow-matching, or the noise prediction for diffusion, toward emotion $e$. Empirically, with $\mathbf{d}_e$ drawn from the hidden geometry of the model and $\lambda$ small, the steered states remain close to the learned manifold, consistent with prior activation steering findings \citep{turner2023actadd,li2023iti}.

\subsection{Mel-Space Guidance via Differentiable Vocoder}
\label{sec:acoustic}
    Hidden state steering in \Cref{eq:latent_guidance} shifts generation at the representation level but does not directly shape the mel spectrogram, which the frozen vocoder $g_{\mathrm{voc}}$ \citep{siuzdak2024vocos} nonlinearly maps to waveform. To target this remaining acoustic gap, we refine the clean mel estimate $\xhat$ at each step by backpropagating the gradient of an external emotion recognizer applied to the waveform through $g_{\mathrm{voc}}$.

    Let $\calL_{\mathrm{emo}}(\xhat, e)$ denotes the recognizer cross-entropy loss for target emotion $e$, and $\nabla_{\xhat}\calL_{\mathrm{emo}}$ its gradient with respect to $\xhat$. We normalize this gradient to unit length and scale by $\|\xhat\|$ to keep the step norm a constant fraction of $\|\xhat\|$ across timesteps, yielding the corrected estimate:
\begin{equation}
  \label{eq:acoustic_guidance}
  \xhat'
  = \xhat
  - \eta\, w(t) \cdot
    \frac{\nabla_{\xhat}\,\calL_{\mathrm{emo}}(\xhat,\, e)}
         {\|\nabla_{\xhat}\,\calL_{\mathrm{emo}}(\xhat,\, e)\|
          + \delta}
    \cdot \|\xhat\|\,,
\end{equation}
    where $\eta$ controls the guidance strength, $\delta$ prevents division by zero, and $w(t)$ is a time dependent schedule whose shape reflects gradient reliability across steps. The gradient is most informative in the middle of the denoising trajectory. At the start steps of the denoising, noise dominates $\xhat$ and the gradient is unreliable. Toward the end steps of the denoising, the spectrogram is nearly converged and resists change. In between, $\xhat$ is both clean enough for reliable gradients and flexible enough to adjust. We therefore concentrate guidance there with a cosine schedule:
\begin{equation}
\label{eq:wschedule}
    w(t) = \tfrac{1}{2}\bigl(1 + \cos\tfrac{\pi\,|t - t_{\mathrm{peak}}|}{t_{\mathrm{width}}}\bigr),
\end{equation}
    which peaks at $t_{\mathrm{peak}}$ and vanishes at distance $t_{\mathrm{width}}$ from the peak, with $w(t) = 0$ taken for $|t - t_{\mathrm{peak}}| > t_{\mathrm{width}}$. To avoid a large guidance step that overshoots the target emotion or introduces artifacts, we complement the temporal scheduling of \Cref{eq:wschedule} with a norm-bounded trust region $\|\xhat' - \xhat\| \le \gamma\,\|\xhat\|$, where $\gamma$ caps the relative step size. If the unconstrained update exceeds this bound, we rescale it to length $\gamma\|\xhat\|$.

\subsection{Unified Dual-Space Update}
\label{sec:combined}
    At each denoising step, hidden space steering and mel-space guidance are applied sequentially under the augmented conditioning $\mathbf{c}^* = (\mathbf{c}, e)$, which combines the speech embedding $\mathbf{c}$ with the target emotion $e$. The steered hidden state $\mathbf{h}^*$ flows through the remaining layers to produce the clean mel estimate $\xhat$, which the guidance then refines into $\xhat'$ of \Cref{eq:acoustic_guidance}. For diffusion samplers, the network $\epsilon_\theta$ predicts the noise component of $\mathbf{x}_t$ and $\bar\alpha_t$ encodes the cumulative signal retained at step $t$. The clean mel estimate at intermediate state $\mathbf{x}_t$ is given by: 
\begin{equation}
\label{eq:dif_step}
  \xhat = \frac{\mathbf{x}_t - \sqrt{1-\bar\alpha_t}\,
  \epsilon_\theta(\mathbf{x}_t, t;\,\mathbf{c}^*)}{\sqrt{\bar\alpha_t}}.
\end{equation}    
    The denoising step then renoises the refined $\xhat'$ to the noise level of step $t-\Delta t$, with $\sqrt{\bar\alpha_{t-\Delta t}}$ setting the signal scale and $\sqrt{1-\bar\alpha_{t-\Delta t}}$ restoring the matching noise component:
\begin{equation}
\label{eq:dif_next}
  \mathbf{x}_{t-\Delta t}
  = \sqrt{\bar\alpha_{t-\Delta t}}\,\xhat'
  + \sqrt{1 - \bar\alpha_{t-\Delta t}}\,
    \epsilon_\theta(\mathbf{x}_t, t;\,\mathbf{c}^*).
\end{equation}

    The same protocol applies to flow-matching samplers with noise prediction replaced by velocity prediction. Combining both interventions into a single per-step update. The velocity prediction gives the clean estimate:
\begin{equation}
  \label{eq:fm_step}
  \xhat = \mathbf{x}_t - t\,v_\theta(\mathbf{x}_t, t;\,\mathbf{c}^*).
\end{equation}
    One backward Euler step with $\xhat'$ then advances $\mathbf{x}_t$ toward $\mathbf{x}_{t-\Delta t}$:
\begin{equation}
\label{eq:fm_next}
  \mathbf{x}_{t-\Delta t} = \mathbf{x}_t + \Delta t \cdot \frac{\xhat' - \mathbf{x}_t}{t}.
\end{equation}

    The $\xhat$ in \Cref{eq:dif_step,eq:fm_step} is derived from the steered predictor at step $t$ and therefore already carries the steered trajectory, exposing a mid-state mel on which guidance can intervene. The refined estimate $\xhat'$ is then propagated to the next step intermediate $\mathbf{x}_{t-\Delta t}$ via \Cref{eq:dif_next,eq:fm_next}, and the protocol repeats until the trajectory reaches $\mathbf{x}_0$, which the vocoder renders into the final waveform $\mathbf{y} = g_{\mathrm{voc}}(\mathbf{x}_0)$. The two interventions cover complementary scopes at each denoising step: steering biases the trajectory along the target emotion direction in hidden space, while guidance refines $\xhat$ in mel space through gradients backpropagated from a waveform recognizer via the vocoder $g_{\mathrm{voc}}$.

\section{Experiments}
\label{sec:experiments}
     We plug DUET into five architecturally diverse pretrained TTS backbones spanning diffusion and flow-matching paradigms and compare with 10 supervised emotional TTS baselines across three datasets, with the comparison validating effectiveness and an ablation study confirming the complementarity of dual-space intervention. We further analyze the representational structure and cross-backbone generalization of our findings. A subjective evaluation and an Ameca humanoid robot deployment further demonstrate perceptually convincing emotional expressiveness and embodied affective interaction enabled by DUET.
\subsection{Experimental Setup}
\label{sec:setup}
    To validate the generalization of DUET, we evaluate on five architecturally diverse frozen TTS backbones spanning with reference speech and text-only conditioning: F5-TTS \citep{chen2025f5tts} with DiT flow-matching, Matcha-TTS \citep{mehta2024matcha} with Transformer flow-matching, GradTTS \citep{popov2021gradtts} with score-based U-Net diffusion, ProDiff \citep{huang2022prodiff} with DDPM and progressive distillation, and StableTTS\footnotemark[4] with lightweight flow-matching. All evaluations use the test splits of ESD \citep{zhou2022emotional}, CREMA-D \citep{cao2014cremad}, and IEMOCAP \citep{busso2008iemocap}, with angry, happy, and sad as the common target emotions across all three datasets.

    We compare DUET with 10 supervised baselines trained with emotion supervision, all evaluated on the same test sets. We use two independently fine-tuned speech emotion recognition (SER) \citep{zhang2026learning} models, HuBERT-large \citep{hsu2021hubert} and WavLM-large \citep{chen2022wavlm}, two architecturally distinct self-supervised speech encoders, adapted per dataset independently, to mitigate evaluator bias from any single architecture. All accuracy values reported in this paper are macro-averaged across these two recognizers. Neither recognizer is used inside the guidance pipeline or for probe labeling, ensuring full independence between evaluation and generation.

    For reproducibility, we extract steering directions once per emotion on the training split and fix them throughout evaluation. The acoustic objective $\calL_{\mathrm{emo}}$ uses emotion2vec \citep{ma2024emotion2vec}, also the probe labeler, distinct from the evaluation recognizers, and guidance strength is set per backbone based on model scale \citep{brack2023sega,ye2024tfg}. Every configuration is evaluated on the full test sets across all three datasets on a single NVIDIA H100 GPU, with all reported numbers are averaged over three independent runs.

\subsection{Plug-and-Play DUET vs Supervised Emotional TTS Baselines}
\label{sec:main_results}

\begin{table}[t]
\centering
\newcommand{\rd}[1]{\textcolor{red!70!black}{#1}}%
    \caption{Comparison of emotion accuracy between DUET plugged into different frozen backbones without retraining and 10 emotional TTS baselines, all trained with emotion supervision on emotional corpora. We report per-class accuracy on angry, happy, and sad, the three emotions common to all datasets, enabling consistent comparison while exposing per-class imbalance. The average accuracy (Avg) summarizes overall performance for ranking. All values are mean accuracy (\%, $\uparrow$) across two independent SER models. \rd{Red}: exceeds the best baseline in that column. $\dagger$: uses a reference speech with neutral emotion at inference.}
\label{tab:main}
\small
\resizebox{\textwidth}{!}{%
\begin{tabular}{l !{\vrule width 0.8pt} ccc|>{\columncolor{gray!6}}c !{\vrule width 0.8pt} ccc|>{\columncolor{gray!6}}c !{\vrule width 0.8pt} ccc|>{\columncolor{gray!6}}c}
\toprule
 & \multicolumn{4}{c}{ESD}
 & \multicolumn{4}{!{\vrule width 0.8pt}c}{CREMA-D}
 & \multicolumn{4}{!{\vrule width 0.8pt}c}{IEMOCAP} \\
\cmidrule(lr){2-5} \cmidrule(lr){6-9} \cmidrule(lr){10-13}
Method
  & Angry & Happy & Sad & Avg
  & Angry & Happy & Sad & Avg
  & Angry & Happy & Sad & Avg \\
\midrule
Qwen3-TTS (2026) \cite{qwen3tts2026} & 27.7 & 71.5 & 41.3 & 46.8
  & 33.9 & 54.6 & 25.4 & 38.0
  & 82.5 & 32.0 & 18.2 & 44.2 \\
CosyVoice2$^\dagger$ (2024) \cite{cosyvoice2_2024} & 21.3 & 43.2 & 59.3 & 41.3
  & 28.5 & 31.3 & 39.2 & 33.0
  & 51.0 & 39.8 & 28.8 & 39.9 \\
EmoVoice (2025) \cite{emovoice2025} & 50.7 & 13.0 & 27.0 & 30.2
  & 32.3 & 0.0 & 20.8 & 17.7
  & 54.2 & 28.4 & 21.5 & 34.7 \\
Chatterbox (2025)\footnotemark[3] & 6.2 & 18.0 & 10.8 & 11.7
  & 4.1 & 13.6 & 36.9 & 18.2
  & 20.1 & 39.0 & 29.9 & 29.7 \\
ChatTTS (2024)\footnotemark[1] & 13.2 & 39.1 & 32.2 & 28.2
  & 0.2 & 23.6 & 26.4 & 16.7
  & 13.2 & 74.6 & 1.4 & 29.7 \\
IndexTTS2 (2025) \cite{indextts2_2025} & 12.7 & 42.2 & 25.7 & 26.9
  & 0.0 & 0.0 & 86.2 & 28.7
  & 0.3 & 54.1 & 46.0 & 33.5 \\
OpenAudio (2025)\footnotemark[2] & 33.5 & 38.8 & 2.8 & 25.0
  & 33.9 & 39.5 & 10.3 & 27.9
  & 68.0 & 16.9 & 1.0 & 28.6 \\
EmoSphere++ (2024) \cite{cho2024emosphere} & 0.0 & 28.6 & 33.9 & 20.8
  & 0.0 & 4.6 & 82.6 & 29.1
  & 0.0 & 42.4 & 18.9 & 20.4 \\
EmotiVoice (2024)\footnotemark[5] & 5.0 & 30.0 & 3.3 & 12.8
  & 21.5 & 10.8 & 17.9 & 16.7
  & 64.7 & 9.5 & 0.5 & 24.9 \\
EmoKnob$^\dagger$ (2024) \cite{emoknob2024} & 25.2 & 81.3 & 16.7 & 41.1
  & 7.4 & 42.6 & 59.0 & 36.3
  & 37.1 & 77.7 & 13.8 & 42.9 \\
\midrule
\multicolumn{13}{l}{\textit{Ours (plug-and-play)}} \\
\rowcolor{gray!10}
\textbf{+ GradTTS (2021)} \cite{popov2021gradtts}
  & \rd{75.0} & 73.8 & \rd{77.8} & \rd{75.5}
  & \rd{89.2} & 23.1 & 68.2 & \rd{60.2}
  & \rd{86.3} & 29.4 & \rd{49.5} & \rd{55.1} \\
\rowcolor{gray!10}
\textbf{+ F5-TTS (2025)}$^\dagger$ \cite{chen2025f5tts}
  & 40.9 & 75.2 & \rd{78.7} & \rd{64.9}
  & \rd{41.8} & 8.7 & \rd{100.0} & \rd{50.2}
  & 11.6 & \rd{83.7} & \rd{97.1} & \rd{64.1} \\
\rowcolor{gray!10}
\textbf{+ Matcha (2024)} \cite{mehta2024matcha}
  & 26.3 & \rd{88.5} & \rd{78.2} & \rd{64.3}
  & \rd{61.8} & 10.8 & 69.8 & \rd{47.4}
  & \rd{88.3} & \rd{83.3} & \rd{55.8} & \rd{75.8} \\
\rowcolor{gray!10}
\textbf{+ ProDiff (2022)} \cite{huang2022prodiff}
  & \rd{53.5} & \rd{89.7} & 47.5 & \rd{63.6}
  & \rd{50.8} & \rd{86.4} & \rd{89.2} & \rd{75.5}
  & 75.2 & \rd{84.7} & 40.7 & \rd{66.9} \\
\rowcolor{gray!10}
\textbf{+ StableTTS (2024)}\footnotemark[4]
  & 29.5 & 47.8 & 52.5 & 43.3
  & 27.9 & \rd{63.3} & 7.2 & 32.8
  & 62.4 & 75.4 & 8.7 & \rd{48.8} \\
\bottomrule
\end{tabular}%
}
\vspace{-12pt}
\end{table}

{\renewcommand{\thefootnote}{}\footnotetext{\scriptsize%
\begin{tabular}[t]{@{}lll@{}}
  $^1$\,\href{https://github.com/2noise/ChatTTS}{github.com/2noise/ChatTTS} &
  $^2$\,\href{https://github.com/fishaudio/open-audio}{github.com/fishaudio/open-audio} &
  $^3$\,\href{https://github.com/resemble-ai/chatterbox}{github.com/resemble-ai/chatterbox} \\
  $^4$\,\href{https://github.com/KdaiP/StableTTS}{github.com/KdaiP/StableTTS} &
  $^5$\,\href{https://github.com/netease-youdao/EmotiVoice}{github.com/netease-youdao/EmotiVoice} &
\end{tabular}}}
  
    We evaluate the effectiveness of DUET plugged into frozen backbones, comparing against 10 supervised state-of-the-art emotional TTS baselines. Four of the five backbones deployed with DUET exceed all supervised baselines in average accuracy. On ESD, DUET plugged into GradTTS achieves the best result of 75.5\%, an absolute improvement of 28.7\% over the strongest baseline Qwen3-TTS at 46.8\%. Even the weakest backbone StableTTS reaches 43.3\%, comparable to the best baselines.

    Baselines exhibit high variance across emotions, with several collapsing on at least one. EmoKnob drops to 16.7\% on sad in ESD and IndexTTS2 to 0\% on angry in CREMA-D, whereas DUET stays balanced across all three emotions. These improvements hold across all three datasets. Results on ESD show the highest absolute accuracy under controlled studio conditions. Results on CREMA-D, which contains 91 speakers, show the largest DUET-baseline margin, indicating robustness to broader speaker variation. Results on IEMOCAP, a conversational corpus, confirm generalization beyond studio reading. Among backbones, GradTTS and ProDiff achieve the strongest results despite their smaller scale, demonstrating that DUET remains effective without requiring large model capacity. StableTTS is the only backbone that does not surpass every supervised baseline, a limitation attributable to its lightweight architecture whose fewer layers constrain its emotion-encoding capacity. Despite this limitation, DUET on StableTTS reaches 48.8\% on IEMOCAP, still surpassing every supervised baseline and confirming the generality of DUET.

\begin{figure}[t]
  \centering
  \subfloat[Per-emotion accuracy gap.\label{fig:per_emo}]{%
    \includegraphics[width=0.32\textwidth]{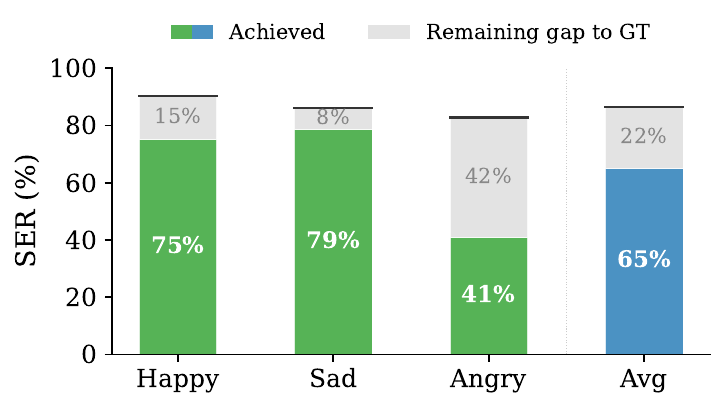}%
  }\hfill
  \subfloat[Orthogonality and direction alignment.\label{fig:repr}]{%
    \includegraphics[width=0.4\textwidth]{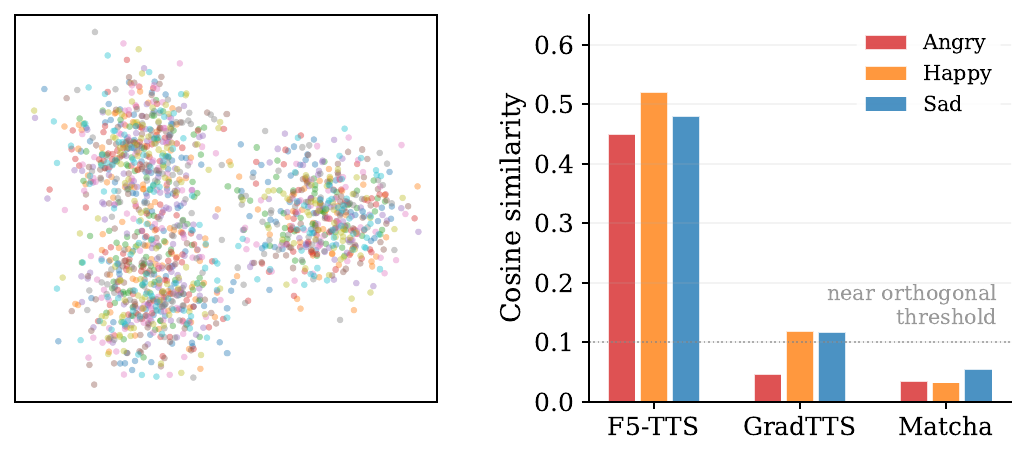}%
  }\hfill
  \subfloat[Probe vs SER accuracy.\label{fig:layer_sweep}]{%
    \includegraphics[width=0.27\textwidth]{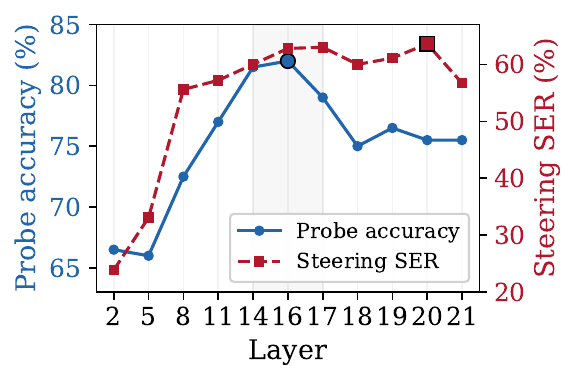}%
  }
  \caption{\textbf{(a)} Colored segments show the accuracy achieved by DUET, the top line marks the GT ceiling of the dataset, and gray segments indicate the gap between them. The noticeably larger gap on angry pinpoints the challenge of capturing its sharp temporal dynamics. \textbf{(b) Left}: scatter panel of hidden states colored by emotion, with speakers distributed uniformly within each cluster, indicating emotion and speaker subspaces are orthogonal. \textbf{(b) Right}: cosine similarity between emotion directions extracted from GT speech and those extracted during generation. For text-only backbones, the two directions are nearly orthogonal. \textbf{(c)} Probe accuracy and steering SER accuracy across layers index for F5-TTS on ESD, closely correlated and both peaking at layer 16.
  }
  \label{fig:exp_main}
  \vspace{-13pt}
\end{figure}

    To investigate why DUET performs differently across emotions, we examine F5-TTS on ESD in \Cref{fig:per_emo}, revealing a consistent asymmetry. Happy and sad reach 83\% and 91\% of the ground truth (GT) ceiling, respectively, while angry reaches only 49\%. We attribute this to anger relying on subtle temporal cues such as sharp onsets and abrupt rhythm changes. Hidden state steering applies a uniform direction across all sequence positions, and mel-space guidance is driven by a speech emotion recognition gradient that reflects whole-utterance emotion. In contrast, happy and sad are characterized by global prosodic patterns such as sustained pitch elevation or flattening, which align with both the uniform hidden perturbation and the utterance-level mel gradient. This asymmetry appears on other backbones as well and motivates temporally adaptive interventions in both spaces for emotions with sharp temporal structure such as anger.
\subsection{Validating the Representation Structure in Hidden States}
\label{sec:repr_analysis}
    We now empirically validate the geometric insight stated in \Cref{sec:method}: that hidden states admit a speaker-coherent geometry for emotion separable from speaker identity, despite no emotion supervision during pretraining. Emotions form well-separated clusters at the discriminative layer $l^*$ in \Cref{fig:discovery_c}, confirming the linear decodability of emotion. The scatter panel in \Cref{fig:repr} further shows speakers uniformly distributed within each emotion cluster, indicating that variations along speaker dimensions span the cluster interior rather than separating emotion classes. The mean cosine similarity between emotion and speaker subspaces is only 0.029, confirming near orthogonality.

    The direction alignment panel in \Cref{fig:repr} further validates the rationale for generation time probing on text-only backbones. The cosine similarities of emotion directions reveal two distinct patterns. For F5-TTS, which accepts reference speech as input, emotion directions extracted with GT speech align well with those extracted during generation. For text-only backbones, the two are nearly orthogonal, with cosine similarities of 0.05 on GradTTS and 0.03 on Matcha. This validates the necessity of generation time probing on text-only backbones, allowing DUET to apply across backbones.

    \Cref{fig:layer_sweep} further validates probe accuracy as the criterion for selecting the intervention layer. It compares probe accuracy with steering SER accuracy obtained when intervention is applied at each layer. The two metrics with similar trends across layers and both peak at layer 16, where steering reaches an SER accuracy of 62.8\%. Early layers 2 to 5 with low probe accuracy produce substantially weaker steering at around 24\% to 33\%. Probe accuracy therefore reliably identifies the most effective intervention layer. Together with the speaker orthogonality and generation time probing results, these detailed analyses validate our findings and further confirm our design in \Cref{sec:method}, which matches the representation structure of pretrained iterative generative TTS models.

\subsection{Ablation of Steering and Guidance}
\label{sec:ablation}
\begin{wraptable}{r}{0.52\textwidth}
  \centering
  \vspace{-12pt}
  \caption{Component ablation of DUET on F5-TTS with ESD. w/o guidance keeps only hidden state steering, and w/o steering keeps only mel-space guidance. Values are SER accuracy (\%, $\uparrow$).}
  \label{tab:ablation}
  \setlength{\tabcolsep}{5pt}
  \begin{tabular}{l|ccc|c}
  \toprule
  Variants & Angry & Happy & Sad & Avg \\
  \midrule
  \textbf{Full}
    & \textbf{40.9} & \textbf{75.2} & \textbf{78.7}
    & \textbf{64.9} \\
  w/o guidance
    & 23.5 & 50.9 & 61.8
    & 45.4 \\
  w/o steering
    & 16.5 & 46.5 & 58.8
    & 40.6 \\
  \bottomrule
  \end{tabular}
  \vspace{-12pt}
\end{wraptable}
    To validate the complementarity of the steering and guidance stages in DUET and the effectiveness of each, we ablate them individually on the F5-TTS backbone. Removing hidden state steering causes a 24.3\% average drop and removing mel-space guidance causes a 19.5\% drop, validating that both components contribute substantially. Per-emotion analysis in \Cref{tab:ablation} shows hidden state steering is the dominant contributor across all three emotions, while mel-space guidance plays a relatively larger role on angry, where spectral gradients can target the transient cues that global steering directions tend to smooth over. This pattern precisely reflects the complementary design of DUET, with hidden state steering handling the global prosodic direction and mel-space guidance refining the transient spectral details. Neither component alone reaches the performance of the full method, confirming that intervention contributes a unique effect that the other cannot fully replicate.

\subsection{Subjective Evaluation}
\label{sec:human_eval}

\begin{wraptable}{r}{0.4\textwidth}
  \vspace{-14pt}
  \centering
  \caption{Subjective evaluation against the three strongest baselines. 20 listeners rate 36 randomly ordered samples on NMOS and EMOS using a 1-5 scale. \textbf{Bold} marks the best results.}
  \label{tab:human}
  \begin{tabular}{l|cc}
  \toprule
  Methods & NMOS$\uparrow$ & EMOS$\uparrow$ \\
  \midrule
  \textbf{DUET} & $3.83$
    & $\mathbf{3.93}$ \\
  Qwen3-TTS & $\mathbf{4.18}$
    & $3.75$ \\
  CosyVoice2 & $4.02$
    & $3.32$ \\
  EmoKnob & $3.54$
    & $3.48$ \\
  \bottomrule
  \end{tabular}
    \vspace{-13pt}
\end{wraptable}
    We complement the objective SER evaluation with a blind subjective evaluation, where 20 listeners rate 36 samples on emotion appropriateness (EMOS) and naturalness (NMOS). As shown in \Cref{tab:human}, DUET achieves the highest EMOS of 3.93, exceeding Qwen3-TTS by 0.18 and CosyVoice2 by 0.61. Its NMOS of 3.83 exceeds EmoKnob at 3.54 despite requiring no explicit emotion supervision, and the moderate gap to Qwen3-TTS reflects the slight spectral perturbation introduced by mel-space guidance. This confirms that emotion control reaches perceptually competitive quality while delivering the highest emotion appropriateness among all tested models.
\begin{figure}[!t]
  \centering
  \includegraphics[width=\textwidth]{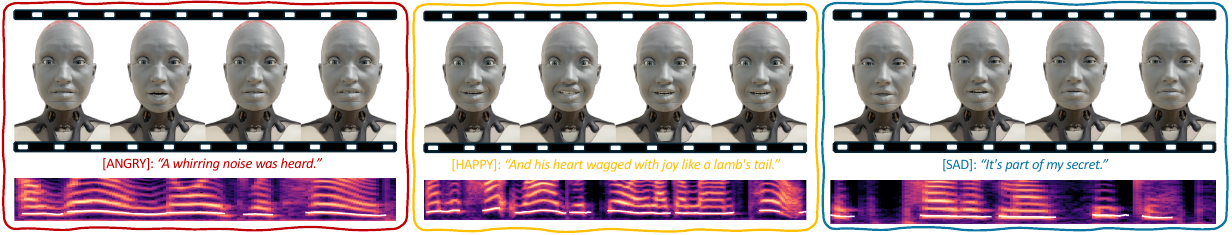}
  \caption{Plug-and-play deployment of DUET on the Ameca humanoid robot. For each of three emotions, \textit{\textcolor[HTML]{DC2626}{ANGRY}}, \textit{\textcolor[HTML]{CA8A04}{HAPPY}}, and \textit{\textcolor[HTML]{2563EB}{SAD}}, the robot speaks the displayed utterance. The filmstrip shows four frames of synchronized speech with an emotion matched facial preset. The mel spectrogram below shows the acoustic signature of the generated audio, with $F0$ contour in white.}
  \label{fig:robot}
  \vspace{-18pt}
\end{figure}
\subsection{Deployment on the Ameca Humanoid Robot}
\label{sec:robot}

    To demonstrate embodied applicability beyond audio-only evaluation, we deploy DUET on an Ameca humanoid robot \citep{cao2025humanoid}, a research platform with lifelike expressions (\Cref{fig:robot}). Speech is generated using DUET plugged into the backbone and streamed to the robot with phoneme-to-viseme lip synchronization and emotion matched facial expression presets. The combination of steered emotional prosody and congruent facial expressions produces coherent cross-modal output. Specifically, the angry utterance pairs raised intensity with a furrowed brow, the happy utterance pairs upward pitch contours with a smile, and the sad utterance pairs slowed cadence with a downcast gaze. Aligning vocal and visual emotion channels is essential for natural human-robot interaction \citep{ugotme2024}, and the Ameca deployment shows that DUET plugs directly into a pretrained TTS model and supplies the speech component of an affective robot.
\vspace{-5pt}
\section{Related Work}
\vspace{-5pt}
\label{sec:related}
\textbf{Inference-time control of generative models.}
    Representation steering and classifier guidance are two independent lines of work for controlling pretrained models. In representation steering, activation addition \citep{turner2023actadd}, inference-time intervention \citep{li2023iti}, representation engineering \citep{zou2025repe}, function vectors \citep{todd2024function}, and low-rank interventions \citep{wu2024reft} shift LLM hidden states along linear concept directions \citep{gurnee2024space}. Representation steering has remained largely confined to language models, where hidden states are static at inference and linear concept geometry is well established \citep{park2024linear, gurnee2024space}. Our analysis confirms this geometry holds for iterative generative TTS hidden states under denoising dynamics, extending representation steering to speech. Classifier guidance \citep{dhariwal2021beatgans} and its training-free generalizations \citep{bansal2023universal,ye2024tfg} steer diffusion sampling via external objectives, with audio applications in Guided-TTS \citep{kim2022guidedtts} and DITTO \citep{novack2024ditto}. These methods do not bridge the mel to waveform gap that determines perceptual quality. DUET combines the two paradigms: hidden state steering at the representation level, and mel-space guidance routing recognizer gradients through the vocoder.

\textbf{Emotion-controllable speech synthesis.}
    Expressiveness in TTS has several lines of work, from learned style embeddings \citep{wang2018gst,skerryryan2018prosody,li2024styletts2} to explicit emotion conditioning via dedicated embeddings or adapters \citep{cho2024emosphere,emoknob2024,guo2023emodiff,tang2023emomix}, typically tied to a single backbone and a fixed emotion taxonomy. Large-scale models such as CosyVoice2 \citep{cosyvoice2_2024}, Qwen3-TTS \citep{qwen3tts2026}, and EmoVoice \citep{emovoice2025} use instruction tuning, requiring massive paired corpora. Training-free EmoSteer-TTS \citep{xie2025emosteer} steers flow-matching TTS via activation vectors and per-position SER selection, and EmoShift \citep{zhou2026emoshift} fine-tunes a per-emotion steering layer. Two gaps remain. First, prior steering has operated on DiT architectures only, inflexible to extending to diffusion-based or other flow-matching backbones. Second, mel-space optimization cannot reliably propagate through the vocoder mapping that produces the actual waveform, and lacks a principled way to verify the resulting correction. DUET addresses the first gap by exploiting a property of pretrained TTS hidden states: emotion is linearly decodable and nearly orthogonal to speaker identity, despite the absence of emotion supervision during pretraining. This geometry, combined with a differentiable vocoder mapping mel to waveform that addresses the second gap, supports a plug-and-play backbone-agnostic framework.
\vspace{-5pt}

\section{Conclusion}
\vspace{-5pt}
\label{sec:conclusion}
    In this paper, we report and analyze the discovery that pretrained iterative generative TTS models, despite never being trained with emotion supervision, encode emotion as a linearly decodable direction in their hidden states, nearly orthogonal to speaker identity. Building on this, we introduced DUET, a plug-and-play framework that achieves emotion control in frozen iterative generative TTS models by unifying hidden state steering and mel-space guidance in a single per-step update. We designed a probing procedure to identify the most discriminative layer and the emotion direction along which the hidden state is steered, shifting the global prosodic trajectory. Mel-space guidance then refines the mel estimate via a gradient routed through a differentiable vocoder, correcting fine spectral details that steering cannot reach. Extensive experiments show that DUET outperforms emotion supervised baselines on both objective and subjective evaluations, validating the generalization and effectiveness of the framework. The results further confirm the value of our discovery, opening a path for emotion control in iterative generative TTS models, with our Ameca deployment demonstrating its plug-and-play applicability in affective robotics.
    
    Two limitations remain. First, the linear steering direction captures global prosodic shifts well but underperforms on emotions whose signal is temporally concentrated, such as anger, because steering applies a uniform direction across all sequence positions, motivating a temporally adaptive variant. Second, DUET is scoped to categorical emotions. The same low dimensional subspace geometry should enable continuous affective control along arousal and valence as a natural extension.

\bibliographystyle{plainnat}
\bibliography{references}

\end{document}